\documentstyle{mn}
\input{epsf} 
\begin{document} 
%version 20.03.98 
   
\def\ee{$e^\pm$}
\def\g{$\gamma$}
% Next lines define "less than or approximately
% equal to", "greater than or approximately equal to", and "approximately
% proportional
\newbox\grsign \setbox\grsign=\hbox{$>$} \newdimen\grdimen %
\grdimen=\ht\grsign
\newbox\simlessbox \newbox\simgreatbox \newbox\simpropbox
\setbox\simgreatbox=\hbox{\raise.5ex\hbox{$>$}\llap
     {\lower.5ex\hbox{$\sim$}}}\ht1=\grdimen\dp1=0pt
\setbox\simlessbox=\hbox{\raise.5ex\hbox{$<$}\llap
     {\lower.5ex\hbox{$\sim$}}}\ht2=\grdimen\dp2=0pt
\setbox\simpropbox=\hbox{\raise.5ex\hbox{$\propto$}\llap
     {\lower.5ex\hbox{$\sim$}}}\ht2=\grdimen\dp2=0pt
\def\simgreat{\mathrel{\copy\simgreatbox}}
\def\simless{\mathrel{\copy\simlessbox}}

\topmargin = -1cm

\title[Discs with Comptonisation and advection]
{Hot accretion discs with thermal Comptonisation and advection
in luminous black hole sources}

\author[A. A. Zdziarski]
{\parbox[]{7in} {Andrzej A. Zdziarski}\\
 N. Copernicus Astronomical Center, Bartycka 18, 00-716 Warsaw, Poland \\ 
}

\date{Accepted 1998 March 18. Received 1997 October 7}

\maketitle

\begin{abstract}
We solve for the structure of a hot accretion disc with unsaturated thermal 
Comptonisation of soft photons and with advection, generalizing the classical 
model of Shapiro et al. The upper limit on the accretion rate due to advection 
constrains the luminosity to $\simless 0.15 y^{3/5} \alpha^{7/5}$ of the 
Eddington limit, where $y$ and $\alpha$ are the Compton and viscosity 
parameters, respectively. The characteristic temperature and Thomson optical 
depth of the inner flow at accretion rates within an order of magnitude of 
that upper limit are $\sim 10^9$ K and $\sim 1$, respectively. The resulting 
spectra are then in close agreement with the X-ray and soft \g-ray spectra 
from black-hole binaries in the hard state and Seyferts. At low accretion 
rates, bremsstrahlung becomes the dominant radiative process. 

\end{abstract}


\section{INTRODUCTION}
\label{s:intro}

Matter accreting onto a compact  object (either in an AGN or a binary) is 
likely to have a net angular momentum, which causes formation of a disc. 
Shakura \& Sunyaev (1973) have developed a theory of optically-thick disc 
accretion. Such discs emit local radiation close to a blackbody, and thus 
their maximal temperature is at most $\sim 10^7$ K. Those discs cannot thus 
explain hard X-rays and soft \g-rays  observed up to $\sim 1$ MeV from both 
black-hole binaries and Seyfert AGNs. Shapiro, Lightman \& Eardley (1976, 
hereafter SLE76) found the existence of another branch of disc solution, in 
which the electron temperature, $T_e$, is $\sim 10^9$ K. In their specific 
model, the dominant radiative process is multiple Compton upscattering of 
copious soft photons by the disc electron plasma, which forms power-law X-ray 
spectra extending up to a few hundred of keV. 

SLE76 assumed in their analysis that all the dissipated energy is locally 
radiated away. However, advection of hot ions is important in a range of 
parameters of the accretion flow (Ichimaru 1977). Solutions including both 
radiative cooling and advection have been extensively studied recently (e.g., 
Abramowicz et al.\ 1995; Narayan \& Yi 1995, hereafter NY95; Chen et al.\ 
1995; Bj\"ornsson et al.\ 1996). Some of those studies (e.g., Bj\"ornsson et 
al.\ 1996) assume Comptonised bremsstrahlung as the only cooling process, 
which turns out to reproduce well the qualitative properties of the 
solutions. However, the predicted X-ray spectra are then very hard, with the 
photon spectral index of $\Gamma\simless 1$, which disagrees with X-ray 
observations of both black-hole binaries and Seyferts, showing typical 
$\Gamma\sim 1.7$--2 (see, e.g, Zdziarski et al.\ 1997, hereafter Z97, for a 
review). Other studies (e.g., NY95, Narayan 1996) include also Comptonisation 
of thermal synchrotron photons in an equipartition magnetic field. However, as 
shown by SLE76, the synchrotron process does not supply sufficient flux of 
soft seed photons to explain the X-ray spectrum of Cyg X-1 in the hard state. 
This conclusion has been extended to GX 339-4 in the hard state and Seyferts 
by Zdziarski et al.\ (1998, hereafter Z98). Those results imply that if 
Comptonised synchrotron emission were the only source of soft photons, the 
X-ray spectra would be harder than those observed. On the other hand, 
soft-photon emission by a cold disc surrounding the hot flow has been shown by 
Esin (1997) either to be negligible or to lead to a runaway cooling of the hot 
flow. 

In spite of these theoretical difficulties, X-ray observations show that a 
supply of soft seed photons sufficient to give rise to spectra with $\Gamma 
\sim 1.7$--2 does exist in luminous black hole systems. Also, the shape of a 
thermal Comptonisation spectrum (at energies much above the seed photon 
energy) is a function solely of $T_e$ and the Thomson optical depth, $\tau$. 
Thus, it is possible to compare a model with X-ray data {\it without\/} 
specifying the soft photon source. Furthermore, most of the power in a 
spectrum with $\Gamma< 2$ is emitted at its high-energy end, and then even the 
model normalization can be compared with data without knowledge of the 
properties of soft seed photons. 

Such an approach was adopted by SLE76, who specified the local cooling rate in 
a disc by assuming a constant value of the Compton parameter, $y$, a quantity 
directly related to the X-ray index $\Gamma$. Observationally, luminous 
black-hole sources typically show an almost constant $\Gamma$ with varying 
flux (e.g., Cyg X-1, Gierli\'nski et al.\ 1997; GX 339-4, Ueda, Ebisawa \& 
Done 1994; IC 4329A, Fiore et al.\ 1992), which justifies that assumption. The 
origin of soft seed photons does not need to be specified in the formalism, 
but SLE76 advocated its origin in cold clouds mixed with the hot flow (e.g.,
Kuncic, Celotti \& Rees 1997). Thermal synchrotron radiation will also provide 
some soft photons. 

In this work, we generalize the hot disc model of SLE76 by including 
advection, by treating the Compton parameter as a free paramater (whereas 
SLE76 assumed $y=1$), and by extending the solution to low accretion rates. As 
we show below, the model predicts $\tau\sim 1$ and $T_e \sim 10^9$ K only 
weakly dependent on the disc radius and the accretion rate. 

Thus, the predictions of the model are in excellent agreement with recent 
studies of broad-band ($\sim 1$--$10^3$ keV) spectra of both black-hole 
binaries in their hard state and Seyferts (e.g., Zdziarski, Johnson \& 
Magdziarz 1996, Z97, Z98; Gierli\'nski et al.\ 1997; Poutanen et al.\ 1997; 
Johnson et al.\ 1997; Magdziarz et al.\ 1998). Those studies show that the 
intrinsic spectra are well fitted by thermal Comptonisation in a plasma with 
just $T_e \sim 10^9$ K and $\tau\sim 1$, with a small dispersion of those 
parameters between different objects. Note that those parameters correspond to 
$y\simless 1$. 

\section{THE HOT DISC SOLUTION}

In the model of SLE76, gravitational energy is converted into ion thermal 
energy, which is then transferred to electrons by Coulomb interactions. 
The electrons, in turn, radiate their energy by Compton upscattering the 
soft seed photons. 

We assume here a pressure equipartition between the magnetic field and gas, 
$P_{\rm mag}= P_{\rm gas}$, which results in $P=2(n_i kT_i+ n_e kT_e)$, where 
$T_i$, $n_i$, $T_e$, $n_e$ are the ion and electron temperature and number 
density, respectively. We then assume the plasma is two-temperature, i.e., 
$T_i\gg T_e$ (which is typically satisfied in an inner part of the disc), 
which allows us to neglect the electron contribution to $P$. The magnetic 
field here is treated similarly to, e.g., NY95, where the viscosity parameter, 
$\alpha$, is unrelated to the magnetic field. In another model, Shakura \& 
Sunyaev (1973) and Sakimoto \& Coroniti (1981) use the magnetic field as a 
source of viscosity, which gives $P_{\rm mag} \approx \alpha P_{\rm gas}$. 
Adopting that model would slightly modify the equations below, slightly 
reducing $\tau$ for a given accretion rate. We further assume \ee\ pairs are 
negligible (as shown to be the case in hot discs with advection by Bj\"ornsson 
et al.\ 1996 and Kusunose \& Mineshige 1996), and neglect the radiation 
pressure. 

We are mostly interested in an inner region of the disc, in which a
self-similar solution of advection-dominated discs found by NY95 breaks down 
(e.g., Chen, Abramowicz \& Lasota 1997). Therefore, we consider here local 
disc solutions and parametrize advection using results of global studies, 
following, e.g., Abramowicz et al.\ (1995) and Bj\"ornsson et al.\ (1996). As 
those authors, we use the pseudo-Newtonian potential (Paczy\'nski \& Wiita 
1980), $\Phi= -GM/(R-R_{\rm S})$, and assume the disc rotation is Keplerian, 
$\Omega=(GM/R)^{1/2}/(R-R_{\rm S})$, where $R$ is the disc radius, $R_{\rm 
S}=2GM/c^2$ is the Schwarzschild radius, and $M$ is the central mass. We take 
into account the presence of He in the flow. 

With the above assumptions, hydrostatic equilibrium and conservation of 
angular momentum yield (with the equation of state of $P= 2n_i kT_i$),
 \begin{equation}
\label{hydro_T_i}
\left(H\over R\right)^2 = {2 r \Theta_i q^2 \over \mu_i},\quad
\Theta_i={J \mu_i \dot m \over 2 \alpha r^{3/2} g q \tau},
\end{equation}
 respectively. Here $H$ is the disc scale height, $r\equiv R c^2/GM$, $q=1-
2/r$, $g=1+4/3rq$, $J=1-(6/r)^{1/2} (3q/2)$ accounts for the zero-torque 
boundary condition at $r=6$, $\Theta_i \equiv kT_i/m_p c^2$, $\alpha$ is the 
viscosity parameter, $\tau=n_e \sigma_{\rm T} H$, $\dot m\equiv \dot M 
c^2/L_{\rm E}$, $\dot M$ is the mass accretion rate, the Eddington luminosity 
is $L_{\rm E}\equiv 4\pi \mu_e GM m_p c/\sigma_{\rm T}$, $\mu_i= 4/(1+3X)$ and 
$\mu_e=2/(1+X)$ are the mean ion and electron molecular weight, respectively, 
$X$ ($\approx 0.7$) is the H mass fraction, and $\sigma_{\rm T}$ is the 
Thomson cross section. 

The energy is dissipated in the disc at a rate of $Q= (3/4\pi) GM\dot M J 
g /q^2 R^3$ per unit area (including both sides of the disc). This energy is 
partly radiated and partly advected to smaller radii. We find that a 
quantity useful to describe the local emission of the disc is a local 
Eddington ratio, defined as the power per unit $\ln r$ in units of $L_{\rm 
E}$, 
$\ell\equiv (dL/d\ln R)/L_{\rm E}= 2\pi Q R^2/ L_{\rm E}$. The dissipative 
$\ell$ is, 
 \begin{equation}
\label{dissipation}
\ell= {3J g\dot m \over 2 q^2 r} \equiv A \dot m\,,
\end{equation} 
 where the function $A$ is needed below. The maximum of $\ell(r)$ occurs 
at $r\approx 16.9$ (compared to $r=27/2$ in the Newtonian, $q=g=1$, $J=1-
6^{1/2}/r^{1/2}$, case). A fraction of $\ell$ is Coulomb-transferred to the 
electrons and then radiated. The rate of Coulomb transfer corresponds to 
(Spitzer 1956), 
 \begin{equation}
\label{Coulomb0}
\ell_{\rm Coul}= {3\over (2\pi)^{1/2}} {m_e\over m_p} {f_{\rm C} \over 
\mu_i} \left( H\over R\right)^{-1} {r \tau^2 \over \Theta_e^{3/2}}
\left( \Theta_i -{m_e\over m_p} \Theta_e\right) \,. \end{equation} 
 Here $\Theta_e\equiv kT_e/m_e c^2$, $f_{\rm C}\approx \ln\Lambda$, $\Lambda 
\approx 0.3 \Theta_e (H/\tau a_0)^{1/2}$, $a_0$ is the Bohr radius, and 
$\Theta_i\ll \Theta_e\ll 1$ (which condition is typically satisfied in a hot 
flow). At this limit, the presence of He reduces the Coulomb rate with respect 
to the case of pure H by $\mu_e/\mu_i$. This is because the rate is 
proportional to $Z^2/A_Z$, which is unity for both H and He, 
and $n_i/n_e=\mu_e/\mu_i$. Here $A_Z$ is the mass number of a $Z$th element. 
When the above condition is not satisfied, the rate given for H by Stepney \& 
Guilbert (1983) can be generalized to yield an $f_{\rm C}$ of, 
 \begin{eqnarray}
\label{Coul_cor}
\lefteqn{ f_{\rm C}  = \ln \Lambda {\pi^{1/2} \Theta_e^{3/2} \over 
2^{1/2} K_2(\Theta_e^{-1}) n_i} \sum_{Z=1} {n_Z Z^2\over A_Z K_2(\Theta_Z^{-
1})}\times}  \nonumber \\ 
\lefteqn{ \left[{2(\Theta_e+\Theta_Z)^2+1\over 
\Theta_e+\Theta_Z} K_1 \! \left({1\over \Theta_e}\! + \! {1\over \Theta_Z}
\right) +2K_0\! \left({1\over \Theta_e} \! +\! {1\over \Theta_Z}
\right) \right],}
 \end{eqnarray}
 where $K_l$ is a modified Bessel function, $n_Z$ is the density of the $Z$th 
element, and $\Theta_Z= \Theta_i/ A_Z$. Note that NY95 neglected the 
effect of $A_Z$ in their accounting for the presence of He in the Coulomb 
rate. Therefore, eq.\ (3.3) in NY95 and eq.\ (A1) in Mahadevan (1997) needs to 
be multiplied by 0.8 and $0.8\mu_e/\mu_i$, respectively, in the limit 
$\Theta_i\ll \Theta_e\ll 1$; otherwise eqs.\ (\ref{Coulomb0})-(\ref{Coul_cor}) 
above should be used.

SLE76 assumed in their solution that a supply of soft seed photons results in 
the Compton parameter (e.g., Rybicki \& Lightman 1979) of $y\equiv 4\Theta_e 
\tau\max(1,\tau)=1$. This can be achieved, e.g., if most of the soft photons 
are from reprocessing of the hard radiation by cold matter in the vicinity of 
the hot flow (Haardt \& Maraschi 1993). A given value of $y$ (which we allow 
here to differ from 1) approximately corresponds to a specific value of the 
energy amplification factor of Comptonisation, $\eta$ (at constant soft photon 
energy, e.g., Dermer, Liang \& Canfield 1991). Then, the local radiative 
balance can be expressed as, 
 \begin{equation}
\ell_{\rm Coul}=\eta(y) \ell_{\rm soft}+\ell_{\rm brem},
\label{eta}
\end{equation}
 where $\ell_{\rm soft}$ and $\ell_{\rm brem}$ corresponds to the power in the 
soft photons and Comptonised bremsstrahlung, respectively. For now, we 
interpret $\ell_{\rm Coul}$ as corresponding to Comptonisation of soft 
photons, which is equivalent to assuming $\ell_{\rm Coul} \gg \ell_{\rm 
brem}$. 
We also assume $T_i\gg T_e$, which allows us to neglect the negative 
term in eq.\ (\ref{Coulomb0}). At the end of this Section, we discuss the 
validity of those assumptions and discuss the behaviour of the solution when 
they break down. 

We assume $\tau\simless 1$ (which we check it is the case {\it a posteriori}), 
which corresponds to $T_e$ and $\ell_{\rm Coul}$ of 
 \begin{eqnarray}
\lefteqn{\Theta_e ={y \over 4\tau},} \label{T_e} \\
\lefteqn{\ell_{\rm Coul}= {2^{3/2} 3 \over \pi^{1/2}} {m_e\over m_p} {f_{\rm 
C} 
J^{1/2}\dot m^{1/2} \tau^3 \over y^{3/2}\alpha^{1/2} q^{3/2} g^{1/2} 
r^{1/4}}\equiv B\dot m^{1/2} \tau^3.} 
 \end{eqnarray} 
 A part of $\ell$ is advected to smaller radii (rather than radiated) at a 
rate of $Q_{\rm adv}=(\dot M/2\pi R^2) (P/\mu_i n_i m_p) \xi$ (Chen \& Taam 
1993), where $\xi$ is a dimensionless factor found $\approx 1$ by global 
hot-disc studies (e.g., Chen et al.\ 1997). The advected power corresponds to, 
 \begin{equation}
\label{advection}
\ell_{\rm adv}= 2 \dot m \Theta_i \xi\mu_i^{-1} = {\xi J \dot m^2 \over \alpha
q g r^{3/2} \tau }\equiv C\dot m^2 \tau^{-1}\,.
\end{equation}

The energy balance of the disc is given by,
 \begin{equation}
\label{balance}
\ell=\ell_{\rm Coul}+\ell_{\rm adv}\,. \end{equation} 
 Hereafter, we assume $\alpha$, $\xi$, $f_{\rm C}$ and $y$ to be constant; 
otherwise the solutions below of eq.\ (\ref{balance}) become implicit. The 
assumption of $y(r)=$ constant corresponds to a fixed ratio of the soft photon 
flux to the total local flux (eq.\ [\ref{eta}]), which can plausibly result 
from physical feedback, related to, e.g., geometry. The solution has two 
branches, corresponding to dominant cooling (SLE76), $\ell\approx \ell_{\rm 
Coul}\propto \dot m$, and dominant advection, $\ell\approx \ell_{\rm adv}$. 
The first one is, 
 \begin{equation} 
\label{sle76} 
\dot m= {2^5 m_e^2\over \pi m_p^2} {f^2_{\rm 
C} q r^{3/2} \tau^6 \over y^{3} \alpha J g^3}, \,\, \ell_{\rm Coul}(r, \tau) 
={2^4 3 m_e^2\over \pi m_p^2} {f^2_{\rm C} r^{1/2} \tau^6 \over y^{3} \alpha q 
g^2} . \end{equation} 
 On the other hand, dominant advection corresponds to, 
 \begin{eqnarray}
\lefteqn{\dot m= {3 \alpha g^2 r^{1/2} \tau\over 2\xi q} , 
\,\, \ell_{\rm Coul}(r,\tau) ={3^{3/2} 2 m_e f_{\rm C} (J g)^{1/2} \tau^{7/2} 
\over \pi^{1/2} m_p y^{3/2} \xi^{1/2} q^2}, \label{adv_dom} } \\ 
\lefteqn{\ell_{\rm Coul}(r,\dot m) ={2^{9/2} m_e \over 3^2 \pi^{1/2} m_p} 
{f_{\rm C} \xi^3 J^{1/2} q^{3/2} \dot m^{7/2} \over \alpha^{7/2} y^{3/2} 
r^{7/4} g^{13/2}} . } \end{eqnarray} 
 Note no dependence on $\alpha$ in $\ell_{\rm Coul}(r,\tau)$. 
 
\begin{figure} 
\begin{center} 
\leavevmode 
\epsfxsize=7.5cm \epsfbox{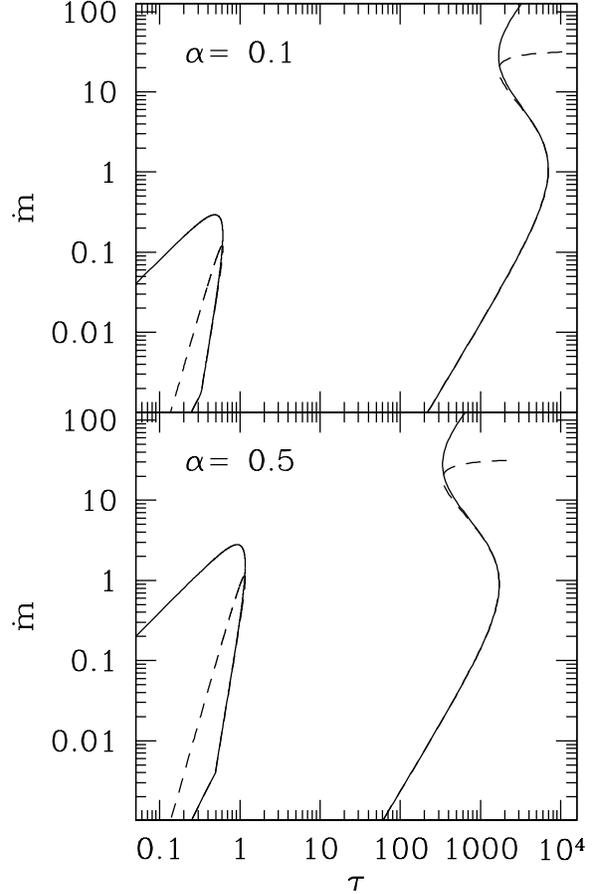} 
\end{center} 
\caption{Disc solutions at $r=27/2$. The left-hand-side curves show the 
optically-thin solution for $\xi=1$, $f_{\rm C}=15$. The disc emission is due 
to Comptonisation with $y=1$ except for the solid curves below the break (at 
$\dot m\sim 0.002$--0.004), for which bremsstrahlung dominates. The 
right-hand-side curves show the optically-thick solution for $M= 10M_\odot$
and with a corona above the disc dissipating 1/2 of the total power (Svensson 
\& Zdziarski 1994). Solid curves give the accreted $\dot m$ whereas the dashed 
curves give $\dot m_{\rm rad}$, i.e., the part of $\dot m$ which is converted 
into escaping radiation. Note a good agreement of the $\dot m$-$\tau$ 
dependence presented here with those of Abramowicz et al.\ (1995), Chen et 
al.\ (1995) and Bj\"ornsson et al.\ (1996). 
 } 
\end{figure} 

The general solution of eq.\ (\ref{balance}) is, 
 \begin{equation} 
\tau = {(A-C a)^{2/5} a^{1/5} \over B^{2/5}} \quad {\rm or}\quad 
\dot m = {(A-C a)^{2/5} a^{6/5} \over B^{2/5}}, \end{equation} 
 where the parameter $a\equiv \dot m/\tau$ is in the range $0<a<3 \alpha g^2 
r^{1/2}/2\xi q$ (with the lower and upper limits corresponding to dominant 
cooling and advection, respectively), and the functions $A$, $B$, $C$ are 
defined in equations above.  The local disc emission corresponds to, 
 \begin{equation}
\ell_{\rm Coul}= (A-Ca) \dot m={(A-Ca)^{7/5} a^{6/5}\over B^{2/5} }.
\end{equation}
 Examples of the solution are shown in Fig.\ 1 (left-hand-side curves). Solid 
curves give $\dot m(\tau)$, and dashed curves give $\dot m_{\rm rad}(\tau)
\equiv \ell_{\rm Coul}/A$, i.e., the locally-radiated fraction of $\dot m$ 
($m_{\rm rad}=\dot m$ on the cooling branch). The corresponding temperature is 
given by eq.\ (\ref{T_e}), $T_e\approx 1.5(y/\tau) \times 10^9$ K. The maxima 
of $\dot m$, $\dot m_{\rm rad}$ and $\tau$ are located in the turning region 
connecting the cooling and advection branches at $a=3A/4C$, $6A/13C$ and 
$A/3C$, respectively. The maximum $\dot m$ is, 
 \begin{equation}
\label{mdot_max}
\dot m_{\rm max}(r) ={3^{12/5} \pi^{1/5}\over 2^{27/5}} \left( m_p\over 
m_e\right)^{2/5} {y^{3/5} \alpha^{7/5} \over f_{\rm C}^{2/5} \xi^{6/5} } 
{J^{1/5} g^3 r^{3/10} \over q^{7/5}}. \end{equation} 
 Numerically, $\dot m_{\rm max} \approx 2.85 y^{3/5} \alpha^{7/5} J^{1/5} g^3 
q^{-7/5} r^{3/10}$, and, e.g., $\dot m_{\rm max}\simeq 7.37 y^{3/5} 
\alpha^{7/5}$ at $r=27/2$. Hereafter, we set $\xi=1$ and $f_{\rm C}=15$ 
(typical for stellar-mass black hole sources; for AGNs, $f_{\rm C}\approx \ln 
\Lambda \approx 20$) in numerical expressions. The $\tau$ corresponding to 
$\dot m_{\rm max}$ is given by
 \begin{equation}
\label{tau_max}
\tau(\dot m_{\rm max})={3^{2/5} \pi^{1/5}\over 2^{12/5} } \left( m_p\over 
m_e\right)^{2/5} {y^{3/5}\alpha^{2/5} \over \xi^{1/5} f^{2/5}_{\rm C} } 
{J^{1/5} g \over q^{2/5} r^{1/5} }. \end{equation} 
 Its maximum of $\tau(\dot m_{\rm max}) \approx 1.22 y^{3/5} \alpha^{2/5}$ is 
at $r=12.9$. Advection at $\dot m_{\rm max}$ carries 0.59 of the dissipated 
energy, and the remainder corresponds to $\ell_{\rm Coul}(\dot m_{\rm max})$ 
of, 
 \begin{equation}
\label{l_max}
\ell_{\rm Coul}(\dot m_{\rm max}) ={3^{17/5} \pi^{1/5}\over 2^{42/5} } \! 
\left( m_p\over m_e\right)^{2/5}\!  {y^{3/5} \alpha^{7/5} J^{6/5} g^4 \over 
f_{\rm C}^{2/5} \xi^{6/5} q^{17/5} r^{7/10} }, \end{equation} 
 or $\ell_{\rm Coul}(\dot m_{\rm max}) \approx 1.07 y^{3/5} \alpha^{7/5} 
J^{6/5} g^4 q^{-17/5} r^{-7/10}$ numerically. Its maximum of $\ell_{\rm 
Coul}(\dot m_{\rm max}) \approx 0.050 y^{3/5} \alpha^{7/5}$ is at $r=18.6$. 
The 
maxima of $\tau$ and $\ell_{\rm Coul}$ in the $\dot m$-$\tau$ space (Fig.\ 1) 
are located in the cooling-dominated region and are larger than the values 
corresponding to $\dot m_{\rm max}$ by the factors of $(4/3)^{4/5}\approx 
1.26$ and $2^{32/5} 7^{7/5}/13^{13/5}\approx 1.63$, respectively. 

We have assumed so far $T_i\gg T_e$, $\ell_{\rm Coul}\gg \ell_{\rm brem}$, 
where 
 \begin{equation}
\label{brem}
\ell_{\rm brem}= {2^{5/2}\over \pi^{3/2}} {m_e\over m_p} \alpha_f
\left( H\over R\right)^{-1} r \tau^2 A_{\rm brem} \Theta_e^{1/2},
\end{equation} 
 $A_{\rm brem}\geq 1$ is a factor including both Comptonisation and 
relativistic corrections and $\alpha_f$ is the fine-structure constant 
(Stepney \& Guilbert 1983). (Note that the bremsstrahlung rate given in eq.\ 
[3.5] in NY95 appears incorrect: 1.25 should be replaced by $\mu_e$.). The 
temperature ratio (at $\tau\simless 1$) and the relative bremsstrahlung rate 
(at $T_e \ll T_i$) are given by, 
 \begin{equation}
\label{T_brem_ratio}
{T_e\over T_i}= {m_e \alpha y\over 2 m_p \mu_i}
{ g q r^{3/2} \over J \dot m},\quad 
{\ell_{\rm brem}\over \ell_{\rm Coul}}={\alpha_f A_{\rm brem} \alpha y^2\over 
3\pi f_{\rm C} \tau} { g q r^{3/2} \over J \dot m}. \end{equation} 
 We see that both ratios have very similar dependences on the disc parameters. 
On the advection branch, $\tau\propto \dot m$, and then $\ell_{\rm brem}/ 
\ell_{\rm Coul}\propto \dot m^{-2}$, which implies bremsstrahlung becoming 
dominant at low $\dot m$ while still $T_e\ll T_i$. However, this happens only 
at very low $\ell_{\rm Coul}$ (below the range of $\dot m_{\rm rad}$ shown in 
Fig.\ 1). On the cooling branch, both above assumptions break down roughly at 
the same time. Note that the effects on $\tau(\dot m)$ of the corrections due 
to $T_e\sim T_i$ and due to bremsstrahlung are in the opposite directions, 
thus partly canceling each other. Numerically, the $T_e\ll T_i$ condition 
(which is marginally more restrictive than $\ell_{\rm brem}\ll \ell_{\rm 
Coul}$) requires $\dot m\gg y\alpha/20$ at $r=27/2$, and $r\ll 280/(\alpha 
y)^{2/3}$ at $\dot m=1$. Otherwise, the disc is dominated by 1-$T$ 
bremsstrahlung, for which the cooling-dominated solution is [for $P=2(n_i 
+n_e) kT$, which modifies eq.\ (\ref{hydro_T_i})], 
 \begin{equation}
\label{mdot_brem}
\dot m= {2^3\alpha_f \over 3\pi^{3/2}} \left[\mu_i \mu_e m_e\over (\mu_i +
\mu_e) m_p\right]^{1/2}\! A_{\rm brem} {q r^{3/2}\over g J}\tau^2.
\end{equation}
From eqs.\ (\ref{hydro_T_i}) and (\ref{T_e}), we also find that a 1-$T$
disc is characterized by a specific local $y$ on both branches when $\tau<1$, 
 \begin{equation}
\label{y1T}
y_{1T}={ 2m_p\mu_e \mu_i J \dot m \over m_e (\mu_e+\mu_i)\alpha g q r^{3/2} }.
\end{equation}
 We see from the above discussion that our main assumptions: $T_e\ll 
T_i$, $\ell_{\rm brem}\ll \ell_{\rm Coul}$, and $y(r)=$ constant, break down 
on the cooling branch roughly simultaneously at low $\dot m$ and/or large $r$. 
Thus, in Fig.\ 1, we show the 2-$T$, constant-$y$, solution at large $\dot m$ 
and the 1-$T$ bremsstrahlung solution at small $\dot m$. In the latter, we set 
$A_{\rm brem}= 1$ since both $y_{1T}\ll 1$ (eq.\ [\ref{y1T}]) and $\Theta_e\ll 
1$ (eq.\ [\ref{hydro_T_i}]) at $\dot m\ll 1$. We simply assume a sharp 
transition between the two regimes; in reality, it will be smooth and more 
complex. We note that at radii at which bremsstrahlung dominates, $\dot m_{\rm 
max}\propto r^{-1/2}$ (Abramowicz et al.\ 1995), whereas $\dot m_{\rm 
max}\propto r^{3/10}$ in the constant-$y$, 2-$T$, region (eq.\ 
[\ref{mdot_max}]). 

In contrast to the above results, Liang \& Thompson (1979) and Wandel \& Liang 
(1991) extended the 2-$T$ solution of SLE76 (eq.\ [\ref{sle76}]) to low values 
of $\dot m$ obtaining a cooling-dominated, 1-$T$ solution with $y=$ constant, 
$\dot m\propto \tau^{-1}$, $\tau>1$. That extension neglects bremsstrahlung; 
however, we find the bremsstrahlung power on that solution exceeds the 
dissipated power, with the ratio of the two $\propto \tau^3$. Thus, that 
solution branch appears not to exist. Instead, the behaviour of the SLE76 
solution at low $\dot m$ is given by a transition to the 1-$T$ bremsstrahlung 
model, as shown above. 

Note that $y$ is necessarily variable in 1-$T$ flows with $\tau<1$ (eq.\ 
[\ref{y1T}]) as well as it becomes large at small $r$, e.g., $y_{1T} \approx 
7\dot m/\alpha$ at $r=27/2$. This represents a strong argument against their 
presence in inner regions of luminous black-hole systems, with constant 
$y<1$ observed (Section 1). 

\section{DISCUSSION}

We have solved the local structure of a hot accretion disc with the main 
cooling process being thermal Comptonisation of soft photons. Some feedback 
processes are assumed to lead to $y=$ constant in a range of radii, which
corresponds to an approximately constant X-ray spectral index, $\Gamma$, of 
emission from each radius. The total spectrum can be then obtained by 
radial integration. The accretion rate of the hot flow can be constant 
throughout the flow, or, plausibly, it can equal the local maximum rate (eq.\ 
[\ref{mdot_max}]). The latter case corresponds to a a two-phase flow, in which 
a cold flow (e.g., in the form of clouds) carries the difference between the 
total $\dot M$ and that corresponding to the maximum $\dot M$ in the hot flow. 
The latter case maximizes the luminosity, 
 \begin{equation}
\label{total_L}
L_{\rm hot}\approx 0.15 y^{3/5} \alpha^{7/5} L_{\rm Edd},
\end{equation}
 which is obtained by radial integration of eq.\ (\ref{l_max}) ($0.5L_{\rm 
hot}$ comes then from $r\simless 40$ and the upper limit of integration is not 
essential as long as it is $\gg 40$). That case also yields the plasma 
parameters only weakly changing with $r$, in particular $\tau\sim 1$ over the 
hot flow (eq.\ [\ref{tau_max}]). Interestingly, the limit of eq.\ 
(\ref{total_L}) appears to agree with the maximum luminosities observed from 
black-hole X-ray novae and persistent X-ray binaries in the hard state (Tanaka 
\& Lewin 1995; Tanaka \& Shibazaki 1996; Phlips et al.\ 1996; Harmon et al.\ 
1994) and AGNs (Peterson 1997) provided $\alpha$ is large (cf.\ Narayan 1996; 
Z98).

The optical depth of the hot flow close to $\dot M_{\rm max}$ is $\tau\sim 1$ 
(eq.\ [\ref{tau_max}]), and $T_e\sim 10^9$ K (eq.\ [\ref{T_e}]), which agree 
well with the observed X-ray spectra (see Section 1). The values of $\tau$ and 
$T_e$ are basically independent of $M$ (except for a slight dependence via the 
Coulomb logarithm), which can explain the close similarity of the broad-band 
spectra of black-hole binaries in the hard state and Seyferts (e.g., Z97, 
Z98). Also, the solutions very weakly depend on the advection parameter, $\xi$ 
[e.g., $\tau(\dot m_{\rm max})\propto\xi^{-1/5}$], which accurate value is 
obtainable only from global studies. In fact, a more important effect is 
geometry of Comptonisation models, which should be specified in comparing fit 
results with hot disc models. E.g., fits in Z97 are for a hemisphere geometry 
(Poutanen \& Svensson 1996), in which case the fitted radial $\tau$ is higher 
than the half-thickness $\tau$ obtained from fitting a slab model. For 
example, Z98 obtains for GX 339-4 $\tau\simeq 2$ and 0.8 in the hemisphere and 
slab geometry, respectively. The latter geometry is more appropriate for hot 
discs radiating close to $\dot M_{\rm max}$, with relatively important 
cooling. 

Black hole systems (in the hard state) show weak spectral variability of the 
spectral shape with changing flux (see Section 1). We can explain this by the 
weak dependence of $\tau$ (determining the spectral shape) on $\ell_{\rm 
Coul}$. This dependence is $\tau\propto \ell_{\rm Coul}^{1/6}$ on the 
cooling-dominated branch, and $\tau\propto \ell_{\rm Coul}^{2/7}$ on the 
advection-dominated branch (eqs.\ [\ref{sle76}]-[\ref{adv_dom}]). Which branch 
corresponds to existing accretion discs can thus be determined by studying the 
X-ray spectral variability. 

The cooling-dominated branch is viscously stable, but unstable thermally 
(Pringle 1976). We note, however, that this thermal instability is different 
and much weaker from the classical bremsstrahlung-cooling instability of 
Pringle, Rees \& Pacholczyk (1973), in which an increase of temperature 
results in a decrease of the cooling rate per unit area. In the model of 
SLE76, an increase of $T_i$ yields an increase of the cooling rate, but the 
corresponding increase of the local heating is faster. The flow becomes stable 
when advection is significant (e.g., Wu 1997). 

Finally, we compare the hot solution with the optically-thick (cold) disc 
solution of Shakura \& Sunyaev (1973) and Abramowicz et al.\ (1988). The 
right-hand-side curves in Fig.\ 1 show the latter solution, but with 0.5 of 
the dissipation occuring in a corona (Svensson \& Zdziarski 1994). The 
solution above $\dot m\sim 1$ (weakly dependent on $\alpha$) has ${\rm d}\dot 
m/{\rm d}\tau<0$ and is both viscously and thermally unstable (Pringle 1981). 
(The flow becomes again stable due to advection of radiation when $L\sim 
L_{\rm E}$, Abramowicz et al.\ 1988, with $L_{\rm E}$ corresponding to $\dot 
m=16$). Thus, the maximum stable $\dot m\sim 1$ of the cold solution in Fig.\ 
1 corresponds to $\sim 5\%$ of $L_{\rm E}$. Notably, this coincides with the 
luminosity of Cyg X-1 in the soft state (Gierli\'nski et al.\ 1998), which 
supports identification of that state with the lower turning point of the cold 
solution. (The presence of a corona, assumed above, is inferred from 
observations of power law tails extending beyond the disc blackbody in the 
soft state of black-hole binaries, Grove et al.\ 1998; Gierli\'nski et al.\ 
1998.) We see in Fig.\ 1 that the $\dot m_{\rm max}$ of the hot solution can 
be, depending on its $\alpha$, either larger or smaller than the maximum 
stable $\dot m$ of the cold flow. Thus, the soft-to-hard state transition in 
X-ray binaries (e.g., Tanaka \& Lewin 1995) can in principle correspond to 
{\it either\/} an increase or decrease of $\dot M$.  

\section*{ACKNOWLEDGEMENTS}

This research has been supported in part by the Polish KBN grants 2P03D01008 
and 2P03D00614. I thank Marek Gierli\'nski for valuable comments on this work 
and help with testing the solutions, and Bo\.zena Czerny for discussions and 
supplying her optically-thick disc code.


\begin{thebibliography}{}

\bibitem[]{} 
Abramowicz M. A., Chen X., Kato S., Lasota J.-P., Regev O., 1995, ApJ, 438, 
L37

\bibitem[]{} 
Abramowicz M. A., Czerny B., Lasota J.-P., Szuszkiewicz E., 1988, ApJ, 332, 
646 

\bibitem[]{} 
Bj\"ornsson G., Abramowicz M. A., Chen X., Lasota J.-P., 1996, ApJ, 467, 99

%\bibitem[]{} 
%Chen X., 1995, MNRAS, 275, 641

\bibitem[]{} 
Chen X., Abramowicz M. A., Lasota J. P., Narayan R., Yi I., 1995, ApJ, 443, 
L61

\bibitem[]{} 
Chen X., Abramowicz M. A., Lasota J. P., 1997, ApJ, 476, 61

\bibitem[]{} 
Chen X., Taam R. E., 1993, ApJ, 412, 254 

\bibitem[]{} 
Dermer C. D., Liang E. P, Canfield E., 1991, ApJ, 369, 410

\bibitem[]{} 
Esin A. A., 1997, ApJ, 482, 400

%\bibitem[]{} 
%Esin A. A., McClintock J. E., Narayan R., 1997, ApJ, 489, 865

%\bibitem[]{} 
%Esin A. A., Narayan R., Ostriker E., Yi I., 1996, ApJ, 465, 312

\bibitem[]{}
Fiore F., Perola G. C., Matsuoka M., Yamauchi M., Piro L., 1992, A\&A, 262, 37 

%\bibitem[]{} 
%Frank J., King A., Raine D., 1992, Accretion Power in Astrophysics, 2nd 
%edition, Cambridge University Press, Cambridge 

\bibitem[]{}
Gierli\'nski M., Zdziarski A. A, Coppi P. S., Poutanen J., Ebisawa K., Johnson 
W. N., 1998, Nucl.\ Phys.\ B, in press 

\bibitem[]{}
Gierli\'nski M., Zdziarski A. A, Done C., Johnson W. N., Ebisawa K., Ueda Y., 
Haardt F., Phlips B. F., 1997, MNRAS, 288, 958 

\bibitem[]{} 
Grove J. E., Johnson W. N., Kroeger R. A., McNaron-Brown K., Skibo J. G., 
1998, ApJ, 499, in press 

\bibitem[]{} 
Haardt F., Maraschi L., 1993, ApJ, 413, 507

\bibitem[]{} 
Harmon B. A., et al., 1994, ApJ, 425, L17

\bibitem[]{} 
Ichimaru S., 1977, ApJ, 214, 840

\bibitem[]{} 
Johnson W. N., McNaron-Brown K., Kurfess J. D., Zdziarski A. A., Magdziarz 
P., Gehrels N., 1997, ApJ, 482, 173

\bibitem[]{}
Kuncic Z., Celotti A., Rees, M.~J., 1997, MNRAS, 284, 717 

\bibitem[]{}
Kusunose M., Mineshige S., 1996, ApJ, 468, 330

\bibitem[]{}
Liang E. P. T., Thompson K. A., 1979, MNRAS, 189, 421

%\bibitem[]{}
%Luo C., Liang E. P., 1994, MNRAS, 266, 386

\bibitem[]{}
Magdziarz P., Blaes O. M., Zdziarski A. A., Johnson W. N., Smith D. A., 1998, 
MNRAS, in press

\bibitem[]{}
Mahadevan R., 1997, ApJ, 477, 585

\bibitem[]{}
Narayan R., 1996, ApJ, 462, 136

\bibitem[]{}
Narayan R., Yi I., 1995, ApJ, 452, 710 (NY95)

%\bibitem[]{}
%Nowak M. A., 1995, PASP, 718, 1207

\bibitem[]{}
Paczy\'nski B., Wiita P. J., 1980, A\&A, 88, 23

%\bibitem[]{}
%Perola G. C., Piro L., 1994, A\&A, 281, 7

\bibitem[]{} 
Peterson B. M., 1997, Active Galactic Nuclei, Cambridge University Press, 
Cambridge 

\bibitem[]{}
Phlips B. F., et al., 1996, ApJ, 465, 907

%\bibitem[]{}
%Piran T., 1978, ApJ, 221, 652

\bibitem[]{}
Poutanen J., Svensson R., 1996, ApJ, 470, 249 

\bibitem[]{}
Poutanen J., Svensson R., Stern B.,  1997, in Winkler C., Courvoisier T., 
Durouchoux P., eds., ESA SP-382, The Transparent Universe, 401

\bibitem[]{}
Pringle J. E., 1976, MNRAS, 177, 65

\bibitem[]{}
Pringle J. E., 1981, ARA\&A, 19, 137

\bibitem[]{}
Pringle J. E., Rees M. J., Pacholczyk A. G., 1973, A\&A, 29, 179

\bibitem[]{} 
Rybicki G. R., Lightman A. P., 1979, Radiative Processes in Astrophysics,
Wiley-Interscience, New York

\bibitem[]{} 
Sakimoto P. J., Coroniti F. V., 1981, ApJ, 247, 19

\bibitem[]{} 
Shakura N. I., Sunyaev R. A., 1973, A\&A, 24, 337

\bibitem[]{} 
Shapiro S. L., Lightman A. P., Eardley D. M., 1976, ApJ, 204, 187 (SLE76)

\bibitem[]{}
Spitzer, L., 1956, Physics of Fully Ionized Gases, Interscience, New York

\bibitem[]{}
Stepney S., Guilbert P. W., 1983, 204, 1269

%\bibitem[]{}
%Stern B. E., Poutanen J., Svensson R., Sikora M., Begelman M. C., 1995, ApJ, 
%449, L13

\bibitem[]{}
Svensson R., Zdziarski A. A., 1994, ApJ, 436, 599

\bibitem[]{}
Tanaka Y., Lewin W. H. G., 1995, in: Lewin W. H. G., van Paradijs J., van den 
Heuvel E. P. J., eds., X-Ray Binaries, Cambridge University Press, Cambridge, 
126

\bibitem[]{}
Tanaka Y., Shibazaki N., 1996, ARA\&A, 34, 607

\bibitem[]{}
Ueda Y., Ebisawa K., Done C., 1994, PASJ, 46, 107

\bibitem[]{} 
Wandel A., Liang E. P., 1991, ApJ, 380, 84

\bibitem[]{} 
Wu X.-B., 1997, MNRAS, 292, 113

%\bibitem[]{} 
%Zdziarski A. A., 1986, ApJ, 303, 94

\bibitem[]{} 
Zdziarski A. A., Johnson W. N., Magdziarz P., 1996, MNRAS, 283, 193

\bibitem[]{} 
Zdziarski A. A., Johnson W. N., Poutanen J., Magdziarz P., Gierli\'nski M., 
1997, in Winkler C., Courvoisier T., Durouchoux P., eds., ESA SP-382, The 
Transparent Universe, 373 (Z97)

\bibitem[]{} 
Zdziarski A. A., Poutanen J., Miko{\l}ajewska J., Gierli\'nski M., Ebisawa 
K., Johnson W. N., 1998, MNRAS, submitted (Z98)

%\bibitem[]{} 
%\.Zycki P. T., Done C., Smith D. A., 1998, ApJ, in press

\end{thebibliography}
\end{document}